\documentclass[a4paper,10pt]{article}
\hoffset= - 2cm
\usepackage{amssymb,float,amsmath,amsfonts,amsthm,cases,cite,bm,latexsym,mathrsfs}
\usepackage[normalem]{ulem}
\usepackage{graphicx,color,epsfig}
\usepackage{wrapfig}
\usepackage[pdfencoding=auto]{hyperref}
\usepackage[dvipsnames]{xcolor}
\usepackage{hyperref}
\usepackage{nicematrix} 
\definecolor{linkcolor}{HTML}{799B03}
\definecolor{urlcolor}{HTML}{799B03}
\hypersetup{pdfstartview=FitH,linkcolor=linkcolor,urlcolor=urlcolor,colorlinks=true}
\newcommand*{\D}{{\rm d}}

\newcommand*{\bpsi}{\bar{\psi}}
\newcommand*{\bH}{\bar{H_2}}
\newcommand*{\cH}{\mathcal{H}}
\newcommand*{\cG}{\mathcal{G}}
\newcommand*{\cF}{\mathcal{F}}
\newcommand*{\TT}{\mathbf{\Theta}}
\newcommand*{\cW}{\mathcal{W}}
\newcommand*{\cD}{\mathcal{D}}
\newcommand*{\cT}{\mathcal{T}}
\def\[{\begin{equation}}
\def\]{\end{equation}}

\begin{document}

\begin{center}
  {\LARGE \bf  Complete stability for spherically symmetric backgrounds in beyond Horndeski theory}

\vspace{10pt}

\vspace{20pt}
S. Mironov$^{a,b}$\footnote{sa.mironov\_1@physics.msu.ru},
V. Volkova$^{a,c}$\footnote{volkova.viktoriya@physics.msu.ru}
\renewcommand*{\thefootnote}{\arabic{footnote}}
\vspace{15pt}

$^a$\textit{Institute for Nuclear Research of the Russian Academy of Sciences, 117312 Moscow, Russia}

$^b$\textit{Institute for Theoretical and Mathematical Physics, MSU, 119991 Moscow, Russia}

$^c$\textit{Department of Particle Physics and Cosmology, Physics Faculty, MSU, 119991 Moscow, Russia}

\end{center}

\vspace{5pt}

\begin{abstract}
We consider a general static, spherically symmetric background in the quadratic beyond Horndeski theory and analyse the behaviour of linear perturbations in both parity odd and parity even sectors. We derive a full set of stability conditions for an arbitrary static, spherically symmetric solution which guarantees absence of ghosts, gradient instabilities, tachyons and superluminal modes in both sectors.

\end{abstract}

\section{Introduction}
\label{sec:intro}

Compact objects have gained considerable attention recently due to significant advances in the experimental field. With more and more observational data available on e.g. black holes and neutron stars mergers our understanding of the underlying gravitational theory gradually evolves to precision. Even though the major part of data from gravitational waves detection is compatible with General Relativity (GR), still existing inconsistencies
and the continuous progress in gaining observational evidence
spur motivation to formulate the 
distinctive features of modified gravity theories.

In the above mentioned context scalar-tensor theories represent a vast group of models with modified gravity
and serve as a popular playground for constructing solutions which describe compact objects like, for instance, black holes.
Nowadays, the most general covariant scalar-tensor theories with $2+1$ degrees of freedom (DOF) are known as degenerate higher order scalar-tensor theories (DHOST), whose Lagrangian 
involves curvature tensors and terms with 
the second derivative of the scalar field (see Refs.~\cite{Langlois1,Langlois2} for review). The most well studied DHOST subclasses are Horndeski theories~\cite{Horndeski:1974wa} (or, equivalently, the generalized Galileon theories~\cite{Deffayet:2009mn}) and beyond Horndeski theories~\cite{Gleyzes:2014dya}, which are related by means of disformal transformation of metric~\cite{Gleyzes:2014qga,BenAchour:2016cay}.
{Thus, various subclasses of DHOST theories, and in particular both Horndeski and beyond Horndeski theories, have been extensively used to model specific solutions describing black holes and wormholes, see e.g. a review in 
Ref.~\cite{KobaRev} and references therein. However, ensuring stability of these solutions w.r.t. small perturbations turned out to be a non-trivial issue.}

In this note we aim to derive the general action governing linear perturbations over an arbitrary static, spherically symmetric background in the quadratic
\footnote{Quadratic here implies that the Lagrangian of the theory involves terms that are quadratic in the second derivatives of the scalar field.} beyond Horndeski theory. 
The main application of such action is stability analysis,
which ensures that dynamical DOFs have {positive energy} and propagate at (sub)luminal speed
over the specific background solutions within theories in question.
In fact, similar analysis of perturbations' behaviour has been earlier considered in different theories belonging to DHOST family, featuring various specific backgrounds which describe compact objects. 
For instance, perturbations about the black hole solution in Horndeski theory were considered in Ref.~\cite{KobayashiOdd} for the odd parity (axial) sector and in Ref.~\cite{KobayashiEven} for parity even (polar) sector. Moreover, the perturbations about black holes were further discussed in the context of more general DHOST theories, see e.g.~\cite{Takahashi:2019oxz,deRham:2019gha,Tomikawa:2021pca,Takahashi:2021bml,Nakashi:2022wdg}, 
where in some cases the scalar field is linearly time-dependent. Unlike the works above in this note we do not require the background geometry to feature any specific solution but we rather keep the background setting in the most general static, spherically symmetric form.
In result, we formulate a full set of stability conditions based on the quadratic action for perturbations, which ensure that any kind of pathologies (i.e. ghosts, gradient instabilities, tachyons and superluminal modes) are absent about an arbitrary static, spherically symmetric background solution.
In particular, the resulting set of stability constraints is applicable to any compact object solution within the theory in question. We focus on the quadratic subclass of beyond Horndeski theory since it is the minimal class of scalar-tensor theories that in principle can admit linearly stable Lorentzian wormhole solutions. The latter follows from the no-go theorems~\cite{nogo1, nogo2} formulated specifically for wormhole solutions and which basically ban stable Lorentzian wormholes in general Horndeski theories.

This work is
sequential to Refs.~\cite{wormRMV1,wormRMV2} where the stability analysis w.r.t.
high energy modes was considered in both parity odd and parity even sectors.
The existing analysis of Ref.~\cite{wormRMV2} cannot claim completeness since it does not address the potential problem with
''slow'' tachyonic modes, which are majorly governed by the mass matrix in the corresponding action for perturbations. It should be noted, however, that the stability constraints w.r.t. tachyonic instabilities for the odd parity modes were derived already in Ref.~\cite{KobayashiOdd}  for the black hole background and got modified for beyond Horndeski case and an arbitrary background in Ref.~\cite{wormRMV1}. However, similar analysis for tachyons for the parity even modes is still missing. We fill this gap in this note.
In this work we present a complete quadratic action for even-parity perturbations which enables one to formulate a full set of stability conditions, addressing the modes of all energies including tachyons within the quadratic beyond Horndeski theory. 
We also ensure that all modes propagate at (sub)luminal speeds by including the corresponding constraints to the stability set.
This full set of stability conditions is particularly useful for construction of specific stable, static and spherically symmetric solutions by means of reverse engineering, where a suitable Lagrangian, featuring the required background solution,
is derived from equations of motion and stability conditions. This bottom-up approach has been adopted for modelling e.g. partially stable Lorentzian wormholes~\cite{wormRMV1,wormRMV2}. 
Let us note, however, that the full set of stability constraints we derive in the even-parity sector is sufficient to claim stability, but it is not the necessary one. The latter is the result of certain restrictive
assumptions, which were made upon formulating the stability conditions. 

The paper is organised as follows. We briefly introduce the theory and the setting in Sec.~\ref{sec:setup} and concentrate on the corresponding linearized theory in Sec.~\ref{sec:linearized}. We revisit the existing results for the odd parity sector in Sec.~\ref{sec:odd}, while in Sec.~\ref{sec:even} we consider the even parity modes in full generality and derive the quadratic action in a gauge invariant form. In Sec.~\ref{sec:K} and~\ref{sec:G} we revise the stability condition w.r.t. ghosts and radial gradient instabilities, respectively, and in Sec.~\ref{sec:QM} we analyse the mass matrix and formulate the new part of stability constraints w.r.t. tachyons.
We conclude in Sec.~\ref{sec:conclusion}.


\section{Spherically symmetric background in beyond Horndeski theory}
\label{sec:setup}
In this section we briefly introduce the theory and notations. In what follows we consider the quadratic subclass of beyond Horndeski theory:
\begin{multline}
\label{eq:lagrangian}
S
= \int \D^4x \sqrt{-g} \left( F(\pi,X)
+ G_4(\pi,X)R \right.\\\left.
+ \left(2 G_{4X}(\pi,X) - 2 F_4 (\pi,X) \; X\right)\left[\left(\Box\pi\right)^2-\pi_{;\mu\nu}\pi^{;\mu\nu}\right] \right.\\\left.
- 2 F_4 (\pi,X) \left[\pi^{,\mu} \pi_{;\mu\nu} \pi^{,\nu}\Box\pi -  \pi^{,\mu} \pi_{;\mu\lambda} \pi^{;\nu\lambda}\pi_{,\nu} \right]\right),
\end{multline}
where $\pi$ is the scalar field,
$X=-\frac12 g^{\mu\nu}\pi_{,\mu}\pi_{,\nu}$,
$\pi_{,\mu}=\partial_\mu\pi$,
$\pi_{;\mu\nu}=\triangledown_\nu\triangledown_\mu\pi$,
$\Box\pi = g^{\mu\nu}\triangledown_\nu\triangledown_\mu\pi$,
$G_{4X}=\partial G_4/\partial X$. The functions $F$ and $G_4$
are characteristic of Horndeski theories, while the
non-vanishing $F_4$ belongs to beyond Horndeski subclass.  

We focus on static, spherically symmetric background solutions described with the following metric:
\[
\label{eq:backgr_metric}
ds^2 = - A(r)\:dt^2 + \frac{dr^2}{B(r)} + J^2(r) \left(d\theta^2 + \sin^2\theta\: d\varphi^2\right),
\]
where the radial coordinate $r$ runs from $-\infty$ to $+\infty$ and the explicit form of function $J(r)$ may vary (e.g. $J(r) = r$ for black holes and $J(r) = \sqrt{1+r}$ for a wormhole profile). In our setting the scalar field $\pi$ is also static, i.e. $\pi=\pi(r)$, and, hence, $X = -B(r) \pi'^2 /2$, where prime denotes the
derivative with respect to $r$.

In what follows we do not specify the form of metric functions $A(r)$, $B(r)$, $J(r)$ and scalar field $\pi(r)$ since we aim to derive generic stability conditions applicable to any static, spherically symmetric backgrounds in beyond Horndeski theory~\eqref{eq:lagrangian}.

\section{Linearized action in a gauge invariant form}
\label{sec:linearized}

Let us consider perturbations about the static, spherically symmetric background~\eqref{eq:backgr_metric}.
We make use of Regge--Wheeler formalism~\cite{ReggeWheeler} and classify metric perturbations as parity odd and parity even types according to their transformation w.r.t. two-dimensional reflection. 
Further expansion of perturbations into series of spherical harmonics $Y_{\ell m}(\theta,\phi)$ enables one to consider modes with different parity, $\ell$ and $m$ separately as the latter do not mix at the linearized level.

As we mentioned above the complete stability analysis for parity odd modes has been carried out in Ref.~\cite{wormRMV1} for the background~\eqref{eq:backgr_metric} in beyond Horndeski theory~\eqref{eq:lagrangian}. We briefly review the existing results for the odd-parity sector in the following Sec.~\ref{sec:odd} and formulate the corresponding set of stability constraints.

The situation in the even-parity sector is somewhat more complicated, since the mass terms in the quadratic action, which govern the tachyon instabilities, were never presented
\footnote{Let us note that corresponding calculations were apparently carried out by the authors of Ref.~\cite{KobayashiEven}  for the case of a black hole, but were not published in the article.}.
In Sec.~\ref{sec:even} we show the results of calculations at the linearized level of perturbations, which 
enable one to formulate the stability criteria, ensuring
that all types of pathologies including tachyons are absent in the even-parity sector. We devise a corresponding set of stability constraints.

\subsection{Parity odd sector}
\label{sec:odd}

In this section we briefly review the results of the stability analysis for perturbations in the parity odd sector about the background~\eqref{eq:backgr_metric} (see e.g. Ref.~\cite{wormRMV1}  for a detailed derivation). 

The odd parity sector involves only metric perturbation.
The quadratic action for the 
only dynamical DOF 
in the odd-parity sector reads
\footnote{In the odd-parity sector we consider $\ell \geq 2$, since the modes with $\ell=0$ do not exist, while the modes with $\ell=1$ are pure gauges. }:
\[
\label{eq:action_odd_final}
\begin{aligned}
S^{(2)}_{odd} = \int \mbox{d}t\:\mbox{d}r\:\sqrt{\frac{A}{B}} J^2
\frac{\ell(\ell+1)}{2(\ell-1)(\ell+2)}\cdot \frac{B}{A}\left[ \frac{\mathcal{H}^2}{A \mathcal{G}} \dot{Q}^2 - \frac{B \mathcal{H}^2}{\mathcal{F}} (Q')^2 -\frac{\ell(\ell+1)}{J^2}\cdot \mathcal{H} Q^2 - V(r) Q^2 \right] \; ,
\end{aligned}
\]
where $Q$ is the master variable (an overdot stands for a time derivative), $A$, $B$, $J$ are metric functions~\eqref{eq:backgr_metric} and
\begin{equation}
\label{eq:cal_FGH}
{\cal F}=2 G_4 ,
\qquad
{\cal H}={\cal G}= 2 \left[G_4 - 2X G_{4X} + 4 X^2 F_4 \right]
\footnote{We have introduced different notations for $\mathcal{H}$ and $\mathcal{G}$ in action~\eqref{eq:action_odd_final}, despite their equality for the quadratic subclass of beyond Horndeski theory~\eqref{eq:lagrangian}, since these coefficients differ as soon as one adds the cubic subclass to the set-up Ref.~\cite{KobayashiOdd, wormRMV1}. } .
\end{equation} 
The first two terms in square brackets of eq.~\eqref{eq:action_odd_final} determine the absence of ghosts and radial gradient instabilities, respectively.
The third term is the 
angular part of the Laplace operator, which governs stability in the angular direction. The "potential" $V(r)$ in eq.~\eqref{eq:action_odd_final} reads
\begin{equation}
\label{eq:Vr}
\begin{aligned}
V(r) &=  \frac{B {\cal H}^2 }{{2\cal F}}
\left[\frac{{\cal F}'}{{\cal F}} \left(2\frac{{\cal H}'}{{\cal H}} -\frac{A'}{A}+\frac{B'}{B}+ 4\frac{J'}{J}\right) -  \frac{{\cal H}'}{{\cal H}} \left( -\frac{A'}{A}+3 \frac{B'}{B}+4 \frac{J'}{J}\right)\right.\\
&\left. - \left(\frac{A'^2}{A^2} -\frac{A'}{A} \frac{B'}{B} - \frac{A''}{A} +\frac{B''}{B} +4 \frac{B'}{B} \frac{J'}{J}-4 \frac{J'^2}{J^2}+ 4 \frac{J''}{J} \right)  - \frac{4}{J^2 B}\cdot \frac{\cal F}{\cal H} \right],
\end{aligned}
\end{equation}
and governs the possible "slow" tachyonic instabilities. 

In order to ensure that the odd-parity sector is free from any type of instabilities it is {sufficient} to satisfy the following set of constraints:
\begin{subequations}
\label{eq:stability_odd}
\begin{align}
\label{eq:stability_G}
\mbox{No ghosts:}\quad &\mathcal{G} > 0, \\
\label{eq:stability_F}
\mbox{No radial gradient instabilities:}\quad &\mathcal{F} > 0, \\
\label{eq:stability_H}
\mbox{No angular gradient instabilities:}\quad &\mathcal{H} > 0, \\
\label{eq:stability_V}
\mbox{No tachyons:}\quad & V(r) > -\frac{6 \mathcal{H}}{J^2},
\end{align}
\end{subequations}
where the constraint~\eqref{eq:stability_V} ensures stability of the modes with low $\ell$. 
Let us note that this constraint for the potential $V(r)$ may not be particularly restrictive~\cite{Trincherini}. 

Another part of constraints comes from the requirement that the odd-parity mode propagates at (sub)luminal speeds in both radial and angular directions. The corresponding squared speeds of sound read:
\[
\label{eq:speed_odd}	
	c_{r,odd}^2 = \frac{\mathcal{G}}{\mathcal{F}}, \quad c_{a,odd}^2 = \frac{\mathcal{G}}{\mathcal{H}},
\]
and give the following (sub)luminality constraints
\[
	\label{eq:subluminality_odd}
 \cG \leq \cF, \qquad \cG \leq \cH.
\]
In the context of stability (and (sub)luminality) analysis the conditions~\eqref{eq:stability_odd} and~\eqref{eq:subluminality_odd} comprise the complete set of stability constraints for the odd-parity sector
and should be considered as constraints on the Lagrangian functions~\eqref{eq:lagrangian}.

\subsection{Parity even sector}
\label{sec:even}

Let us turn to the parity even modes.
We adopt the following parametrization of metric perturbations $h_{\mu\nu}$ and scalar field perturbation $\delta \pi$ in the parity even sector~\cite{ReggeWheeler}
\[
\label{eq:even_parity}
\mbox{Parity~even}\quad \begin{cases}
\begin{aligned}
h_{tt}=&A(r)\sum_{\ell, m}H_{0,\ell m}(t,r)Y_{\ell m}(\theta,\varphi), \\
h_{tr}=&\sum_{\ell, m}H_{1,\ell m}(t,r)Y_{\ell m}(\theta,\varphi),\\
h_{rr}=&\frac{1}{B(r)}\sum_{\ell, m}H_{2,\ell m}(t,r)Y_{\ell m}(\theta,\varphi),\\
h_{ta}=&\sum_{\ell, m}\beta_{\ell m}(t,r)\partial_{a}Y_{\ell m}(\theta,\varphi), \\
h_{ra}=&\sum_{\ell, m}\alpha_{\ell m}(t,r)\partial_{a}Y_{\ell m}(\theta,\varphi), \\
h_{ab}=&\sum_{\ell, m} K_{\ell m}(t,r) g_{ab} Y_{\ell m}(\theta,\varphi)+\sum_{\ell, m} G_{\ell m}(t,r) \nabla_a \nabla_b Y_{\ell m}(\theta,\varphi)\,\\
\delta \pi=& \sum_{\ell, m}\chi_{\ell m}(t,r)Y_{\ell m}(\theta,\varphi),
\end{aligned}
\end{cases}
\]
where $a,b=\theta, \varphi$ and $g_{ab} = \mbox{diag}(1, \;\sin^2\theta)$, while 
$\nabla_a$ is a covariant derivative on a 2-sphere. 
Since modes with different $\ell$ and $m$ evolve independently at the linear level we consider them separately and drop all related subscripts in what follows.

Then the quadratic action for perturbations~\eqref{eq:even_parity} following from eq.~\eqref{eq:lagrangian} reads (for $\ell \geq 2$)
\footnote{The case of lower momenta $\ell=0,1$ should be considered separately as perturbation parametrization~\eqref{eq:even_parity} significantly simplifies and, hence, derivation of the final quadratic action for dynamical DOFs differs. However, both monopole and dipole cases do not usually give any additional constraints stability-wise~\cite{KobayashiEven,wormRMV1}.}: 
\begin{equation}
\label{eq:action_even}
\begin{aligned}
&S_{even}^{(2)} = \int \mbox{d}t\:\mbox{d}r 
\Big(H_0 \left[ a_1 \chi''+ a_2\chi'+a_3 H_2'+j^2 a_4 \alpha'+\left( a_5+j^2 a_6 \right) \chi
+\left( a_7+j^2 a_8 \right)H_2
\right.  \\ &\left.
+j^2a_9 \alpha +j^2 a_{10} G'' + j^2 a_{11} G' + j^2 a_{12} G + a_{13} K'' +a_{14} K' + (a_{15} + j^2 a_{16}) K
\right]
+j^2 b_1 H_1^2 \\
&+H_1 \left[b_2 {\dot {\chi}}'+b_3 {\dot {\chi}}+b_4 {\dot H_2}+j^2b_5 {\dot \alpha} + j^2 b_{6}\dot{G}' + j^2 b_{7}\dot{G} + b_{8}\dot{K}' + b_{9}\dot{K} + j^2 b_{10}\beta' + j^2 b_{11}\beta\right] \\
&+ {\dot H_2} \left[c_1{\dot {\chi}} + j^2 c_{7} \dot{G} + c_{8}\dot{K}\right] + c_6 H_2^2
+ H_2 \big[c_2  \chi'+\left( c_3+j^2 c_4\right)\chi + j^2 c_5 \alpha 
+j^2 c_{9} G' 
\\ &
+ j^2 c_{10} G + c_{11} K' + (c_{12} +j^2 c_{13})K + j^2 c_{14} \dot{\beta}
\big] + j^2d_1 {\dot \alpha}^2 +j^2 d_4 \alpha^2
+\alpha \big[ j^2 d_2 \chi'+j^2 d_3 \chi
\\&
 + j^2 d_{5} G' + j^2 d_{6} G + j^2 d_{7} K' + j^2 d_{8} \dot{\beta}' + j^2 d_{9} \dot{\beta}\big]
+ e_1{\dot {\chi}}^2+e_2 \chi'^2+\left( e_3+j^2 e_4  \right) \chi^2 
\\&
+ \dot{\chi} \; \big[j^2  e_{5} \dot{G}' + j^2 e_{6} \dot{G}+ e_{7} \dot{K}'  + e_{8} \dot{K} + j^2 e_{9} \beta' + j^2 e_{10}\beta \big]
+ \chi' \;[j^2 e_{11} G' + j^2 e_{12} G + e_{13} K' 
\\&
+ (e_{14}+j^2 e_{15}) K] + \chi \; [j^2 e_{16} G + (e_{17} +j^2 e_{18})K]
+ j^2 p_{1} (\beta')^2 + j^2 p_{2} \beta^2 
+ j^2 \beta \; [ p_{3}\dot{G} +  p_{4}\dot{K} ]
\\&
+ j^2 p_{5} \dot{G}^2 + j^2 p_{6}(G')^2 + j^2 p_{7}G^2 + p_{8}\dot{K}^2 + p_{9}(K')^2 
+ j^2 p_{10}\dot{G}\dot{K} + j^2 p_{11} G'K' + j^2 p_{12} G K'
\Big),
\end{aligned}
\end{equation}
where $j = \ell(\ell+1)$ and coefficients $a_i$, $b_i$, $c_i$, $d_i$, $e_i$ and $p_i$ depend on a background~\eqref{eq:backgr_metric} and the Lagrangian functions and are explicitly given in Appendix A.

The quadratic action~\eqref{eq:action_even} is invariant under infinitesimal 
coordinate change $x^{\mu} \to x^{\mu} + \xi^{\mu}$
where $\xi^{\mu}$ is parametrized as follows
\[
\label{eq:xi}
\xi^{\mu} = \Big(T_{\ell m}(t,r), R_{\ell m}(t,r), \Theta_{\ell m}(t,r) \partial_{\theta}, \dfrac{\Theta_{\ell m}(t,r) \partial_{\varphi}}{\sin^2\theta}\Big)\,  Y_{\ell m}(\theta,\varphi ),
\]
with arbitrary functions $T(t,r)$, $\Theta(t,r)$ and $R(t,r)$. 
The perturbation variables~\eqref{eq:even_parity} obey the 
following gauge transformation laws:
\begin{equation}
\begin{aligned}
\label{eq:gauge_laws}
H_0 &\rightarrow H_0 + \dfrac{2}{A} \dot{T} - \dfrac{A'}{A} B R, 
\qquad \qquad & 
\alpha &\rightarrow \alpha + R + \Theta' - 2 \dfrac{J'}{J} \Theta,
\\
H_1 &\rightarrow H_1 + \dot{R} + T' - \dfrac{A'}{A} T, 
\qquad \qquad & 
K &\rightarrow K + 2 B \dfrac{J'}{J} R,
\\
H_2 &\rightarrow H_2 + 2 B R' + B' R, 
\qquad \qquad & 
G &\rightarrow G + 2 \Theta,
\\
\beta &\rightarrow \beta + T + \dot{\Theta}, 
\qquad \qquad & 
\chi &\rightarrow \chi + B \pi' R,
\end{aligned}
\end{equation}
where we again dropped all the subscripts $\ell, m$ for brevity. Hence, 
choosing specific functions $T$, $\Theta$ and $R$ it is possible to make different sets of variables to vanish. For now we leave the action~\eqref{eq:action_even} in the gauge invariant form. 

Variation of the action~\eqref{eq:action_even}
w.r.t. $H_0$ and $H_1$ gives the following constraint equations,
respectively
\begin{equation}
\begin{aligned}
\label{eq:H0}
\delta H_0: \qquad &a_1 \chi''+a_2  \chi'+a_3 H_2'+
j^2 a_4 \alpha'+\left( a_5+j^2 a_6 \right) \chi
+\left( a_7+j^2 a_8 \right)H_2+j^2a_9 \alpha \\
&+ j^2(a_{10} G'' + a_{11}G' 
+ a_{12}G) +a_{13}K'' + a_{14}K' +(a_{15}+j^2 a_{16})K = 0,
\end{aligned}
\end{equation}
\begin{equation}
\begin{aligned}
\label{eq:H1}
 \delta H_1: \qquad H_1 =& - \frac{1}{2 j^2 b_1} \left( b_2 {\dot {\chi}}'+b_3 {\dot {\chi}}+b_4 {\dot H_2}+j^2b_5 {\dot \alpha} \right. \\
 &\left.+ j^2 b_6 \dot{G}' + j^2 b_7 \dot{G} +b_8 \dot{K}' +b_9 \dot{K} 
 +j^2 b_{10} \beta' +j^2 b_{11}\beta \right).
\end{aligned}
\end{equation}
We note that $\beta$ is also non-dynamical according to the action~\eqref{eq:action_even}, however, variation w.r.t. $\beta$ gives an equation which is a linear combination of eqs.~\eqref{eq:H0} and~\eqref{eq:H1}, so we do not give it here.
In what follows we aim to make use of constraint equations and integrate out the non-dynamical variables, so that we end up with the quadratic action featuring two dynamical DOFs.

Let us now partially fix the gauge by making use of $T$ and $\Theta$ in eq.~\eqref{eq:gauge_laws} so that
\begin{equation}
\label{eq:gauge_G_beta}
G = \beta = 0,
\end{equation}
while we keep the gauge freedom related to function $R$ for now. 
Instead of solving the differential equation~\eqref{eq:H0} we put forward a change of variable, which transforms eq.~\eqref{eq:H0} into the algebraic one w.r.t. $K$.
Hence, we introduce a new variable $\psi$ as follows:
\begin{equation}
\label{eq:alpha_shifted}
\alpha = \psi - \dfrac{1}{j^2 a_4} \left( a_1 \chi' + a_{13} K' \right),
\end{equation}
Then upon substituting $\alpha$ and $\alpha'$ in eq.~\eqref{eq:H0}
both terms with $\chi''$ and $K''$ get cancelled out (explicit form the coefficients involved in the equations can be read off in Appendix A). Moreover, $K'$ also vanishes, so the resulting equation is indeed an algebraic one in terms of $K$ and reads:
\begin{equation}
\begin{aligned}
\label{eq:K}
K = - \dfrac{1}{a_{15} +j^2 a_{16}} &\left[ a_3 H_2' +(a_7+j^2 a_8) H_2 +j^2 a_4 \psi' +j^2 a_9 \psi \right.\\ &\left.+ \left( \dfrac{a_4(a_2 - a_1')+a_1(a_4'-a_9)}{a_4}\right)\chi'  +(a_5+j^2 a_6)\chi.
\right]
\end{aligned}
\end{equation}
So $K$ is expressed in terms of $H_2$, $\psi$ and $\chi$.
Next, by substituting $\alpha$ from eq.~\eqref{eq:alpha_shifted} and $K$ from eq.~\eqref{eq:K} into the second constraint~\eqref{eq:H1} one can express $H_1$ in terms of $H_2$, $\psi$ and $\chi$ as well. Therefore, by making use of both constraint equations~\eqref{eq:H0} and \eqref{eq:H1} and a simple field redefinition~\eqref{eq:alpha_shifted} the quadratic action~\eqref{eq:action_even} can be cast in terms of $H_2$, $\psi$ and $\chi$. 

Finally, we eliminate $\chi$ by fixing the residual gauge 
freedom associated with function $R$
in eq.~\eqref{eq:gauge_laws} and set $\chi = 0$ (Regge-Wheeler-unitary gauge), so we arrive to the following quadratic action in terms of two DOFs $H_2$ and $\psi$:
\begin{equation}
\label{eq:unconstrained_action}
\begin{aligned}
S_{even}^{(2)} = &\int \mbox{d}t\:\mbox{d}r \left(
g_{4}\dot{\psi}'\:^2 + g_{5}\psi''\:^2 
+f_{4}\dot{H_2}'^2  + f_{5} H_2''^2
+ h_{7} \dot{H_2}' \dot{\psi}' +  h_{8} H_2'' \psi''
\right.\\ 
&\left.
+g_{1}\dot{\psi}^2 + g_{2}\psi'\:^2 +  g_{3}\psi^2
 + f_{1}\dot{H_2}^2 + f_{2}H_2'^2 + f_{3}H_2^2 
+ h_{5}\dot{H_2} \dot{\psi}' 
\right.\\ 
&\left.
+ h_{6} H_2' \psi'' + h_{1}\dot{H_2} \dot{\psi}  + h_{2} H_2' \psi' + h_{3} H_2 \psi'  + h_{4} H_2 \psi
\right),
\end{aligned}
\end{equation}
where $g_{j}$, $f_{j}$ and $h_{j}$ are some combinations of the initial coefficients $a_i$, $b_i$, $c_i$, $d_i$ and $p_i$ in eq.~\eqref{eq:action_even} and whose explicit form is irrelevant at this point. 
We see that, at least naively, the linearized equations for both $H_2$ and $\psi$ involve terms with four derivatives (see the first line of eq.~\eqref{eq:unconstrained_action}). 
However, the detailed analysis of the relations between coefficients 
$g_{j}$, $f_{j}$ and $h_{j}$ in Ref.~\cite{wormRMV2} showed that upon field redefinitions
\begin{equation}
\label{eq:newPsiH2}
\begin{aligned}
& \psi = \bpsi + \frac{(2\cH J J' + \Xi \pi')}{2\ell(\ell+1) \; \cH} \bH, \\
& H_2 = \bH + 4 \ell(\ell+1) B J \dfrac{\;{\cal H} }{\TT} \; \bar\psi',
\end{aligned}
\end{equation}
all the terms with more than two derivatives vanish 
in eq.~\eqref{eq:unconstrained_action} 
(both $\Xi$ and $\TT$ are some combinations of Lagrangian functions and background values of metric and scalar field, see eqs.~\eqref{eq:Xi} and~\eqref{eq:Theta} below for explicit definitions)
\footnote{The potential issue with redefinitions~\eqref{eq:newPsiH2} related to their divergence at $\TT=0$ was addressed in detail in Ref.~\cite{wormRMV2}. It was explicitly shown that the solutions to the corresponding linearized equations for the initial ${H}_2$ and ${\psi}$ are regular in the vicinity of $\TT=0$, hence, it was proved that $\TT=0$ is not a special point fro the initial system.}. 
Hence, we arrive to the conventional form of the quadratic action for $2$ DOFs in the even parity sector:
%
%
\begin{equation}
\label{eq:action_KGQM_new}
S_{even}^{(2)} = \int \mbox{d}t\:\mbox{d}r 
\left(
\frac12 {\mathcal{K}}_{ij} \dot{v}^i \dot{v}^j - \frac12 {\mathcal{G}}_{ij} v^{i\prime} v^{j\prime} - {\mathcal{Q}}_{ij} v^i v^{j\prime} - \frac12 {\mathcal{M}}_{ij} v^i v^j
\right),
\end{equation}
where $i=1,2$ with $v^1 = \bar{H_2}$, $v^2 = \bpsi$ and both DOFs are described by the second order differential equations. 
Matrices ${\mathcal{K}_{ij}}$, ${\mathcal{G}_{ij}}$, ${\mathcal{Q}}_{ij}$ and ${\mathcal{M}}_{ij}$ are the 
central objects for stability analysis for any static, spherically symmetric solution~\eqref{eq:backgr_metric}, {since they define both the  energy density and the dispersion relation for perturbations. 
Let us discuss the matrices ${\mathcal{K}_{ij}}$, ${\mathcal{G}_{ij}}$, ${\mathcal{Q}}_{ij}$ and ${\mathcal{M}}_{ij}$ one by one and formulate 
the set of stability constraints for them, which would ensure positivity of energy density and regular dispersion relation for all perturbation modes.} In full analogy with the parity odd sector we aim to derive the constraints on the Lagrangian functions~\eqref{eq:lagrangian}, which would ensure the stability in the even parity sector as well. 

\subsubsection{Matrix ${\mathcal{K}}$: no ghost condition}
\label{sec:K}
Positive definiteness of matrix ${\mathcal{K}_{ij}}$ determines the absence of a ghost about the solution:
\[
\label{eq:noghost}
	\mathcal{K}_{11} > 0, \qquad \det{\mathcal{K}} > 0,
\]
where 
{
\begin{equation}
\label{eq:k11detk_RW}
{\mathcal{K}}_{11} = 
\dfrac{ 1}{4 \ell(\ell+1)(\ell +2)(\ell-1) \sqrt{AB}}
\dfrac{\TT^2}{\mathcal{F}},
\quad
\det{\mathcal{K}} = \dfrac{\ell(\ell+1) B}{4(\ell+2)(\ell-1)A}
\dfrac{(2\mathcal{H} J J' + \Xi \pi')^2 (2 \mathcal{P}_1 - \mathcal{F})}{\mathcal{F}},
\end{equation}
and
\[
\label{eq:Xi}
\Xi =  2G_{4\pi}J^2 + 4G_{4X}BJJ'\pi'
 -2G_{4\pi X}BJ^2\pi'^2
 - 4G_{4XX}B^2JJ'\pi'^3 
   + 16F_{4}B^2JJ'\pi'^3 - 4F_{4X}B^3JJ'\pi'^5,
\]
\begin{subequations}
\begin{align}
\label{eq:Theta}
\TT &= 2 \ell (\ell+1) \: J\left(\mathcal{H} - 2F_4 B^2 \pi'^4\right) +\frac{B(A' J - 2 A J')}{A}
\left(2{\cal H} J J' + \Xi \pi'\right),\\
\label{eq:P1}
\mathcal{P}_1 &= \frac{\sqrt{B}}{\sqrt{A}} \cdot
\frac{\mbox{d}}{\mbox{d}r}\left[\frac{\sqrt{A}}{\sqrt{B}}
~\xi
\right],\\
\label{eq:xi}
\xi &= \frac{J^2 \mathcal{H}\left(\mathcal{H} - 2 F_4 B^2 \pi'^4\right)}{2{\cal H} J J' + \Xi \pi'}
\; ,
\end{align}
\end{subequations}
while $\cH$ and $\cF$ are given in eq.~\eqref{eq:cal_FGH}. It immediately follows from
eq.~\eqref{eq:k11detk_RW} that both inequalities hold provided that:
\[
\label{eq:noghost_explicit}
2 \mathcal{P}_1 - \cF > 0,
\] 
and $\cF > 0$, where the latter is already ensured
with the no radial gradient instability condition~\eqref{eq:stability_F} in the odd-parity sector.

The rest of components of the matrix ${\mathcal{K}_{ij}}$ are given in Appendix B for completeness.

\subsubsection{Matrix ${\mathcal{G}}$: no radial gradient instabilities condition}
\label{sec:G}
To avoid gradient instabilities propagating in the radial direction one has to require positive definiteness of matrix 
${\mathcal{G}_{ij}}$:
\begin{equation}
\label{eq:noradial}
\mathcal{G}_{11} > 0, \qquad \det{\mathcal{G}} > 0,
\end{equation}
with
\begin{equation}
\label{eq:g11detg_RW}
{\mathcal{G}}_{11} = \dfrac{\sqrt{AB} }{4 \ell(\ell+1)(\ell +2)(\ell-1)} \dfrac{\mathcal{H}\TT^2}{\mathcal{F}^2}, \quad
\det{\mathcal{G}}=\dfrac{\ell (\ell+1) A B^3}
{ (\ell+2) (\ell-1)}
\dfrac{\mathcal{G}  \mathcal{N}}{ \mathcal{F}^2 },
\end{equation}}
where
\[\label{eq:N}
	\mathcal{N} = 2 J^2 \Gamma \mathcal{H} \Xi \pi'^2  - \mathcal{G} \Xi^2 \pi'^2-4 J^4 \Sigma \mathcal{H}^2/B,
\]
and
\begin{subequations}
\label{eq:GammaSigma}
\begin{align}
\label{eq:Gamma}
& \Gamma = \Gamma_1 + \frac{A'}{A} \Gamma_2,
\\
& \Gamma_{1} = 4\left(G_{4\pi} + 2XG_{4\pi X} + \frac{B\pi'J'}{J}(G_{4X} + 2XG_{4XX}) \right) - \frac{16J'}{J}B\pi'X(2F_{4} + XF_{4X}), \\
&\Gamma_{2} = 2B\pi'\left(G_{4X} - B\pi'^2G_{4XX}\right)
- 8B\pi'X(2F_{4} + XF_{4X}),\\
\label{eq:Sigma}
& \Sigma = XF_{X} + 2F_{XX}X^2 +2\left(\frac{1 - BJ'^2}{J^2} - \frac{BJ'}{J} \frac{A'}{A}\right)X(G_{4X} + 2XG_{4XX})
\nonumber\\& 
     - \frac{4BJ'}{J}\left( \frac{J'}{J } + \frac{A'}{A}\right)X^2(3G_{4XX} + 2XG_{4XXX}) +  2B\pi'\left(\frac{4J'}{J } + \frac{ A'}{A} \right)X( \frac{3}{2} G_{4\pi X} + XG_{4 \pi XX})
      \nonumber\\& 
      + \frac{B^3J'(A'J + AJ')}{AJ^2}\pi'^4(12F_{4} - 9F_{4X}B\pi'^2 + F_{4XX}B^2\pi'^4).
\end{align}
\end{subequations}
The rest of components ${\mathcal{G}_{ij}}$ are given in Appendix B.
Recalling that $\cG>0$ and $\cH>0$ according to stability constraints in the odd-parity sector~\eqref{eq:stability_G} and~\eqref{eq:stability_H}, in order 
to satisfy the conditions~\eqref{eq:g11detg_RW} one has to additionally require
\[
	\label{eq:noradial_explicit}
\mathcal{N} > 0.
\]

Let us note that together the matrices $\mathcal{K}_{ij}$ and $\mathcal{G}_{ij}$ define the speeds of perturbations propagating in the radial direction, which are given by eigenvalues of matrix $(AB)^{-1}(\mathcal{K})^{-1}\mathcal{G}$ and read:
\begin{equation}
\label{eq:speed_even_radial}
c_{r1}^2 = \frac{\mathcal{G}}{\mathcal{F}},
\qquad
c_{r2}^2 = \frac{\mathcal{N}}
{\left(2{\cal H} J J' + \Xi \pi'\right)^2 (2{\cal P}_1-{\cal F})} \; ,
\end{equation}
where $c_{r1}^2$ coincides 
with the $c_{r,odd}^2$ in eq.~\eqref{eq:speed_odd} and can be interpreted as a speed of gravitational waves.
The requirement of (sub)luminal propagation of both modes gives 
additional constraints for the model:
\[
	\label{eq:subluminal_radial}
	\cG \leq \cF, \qquad \mathcal{N} \leq \left(2{\cal H} J J' + \Xi \pi'\right)^2 (2{\cal P}_1-{\cal F}).
\]

\subsubsection{Matrices ${\mathcal{Q}}$ and ${\mathcal{M}}$: no tachyon conditions and no angular gradient instabilities} 
\label{sec:QM}
Let us turn to the matrices ${\mathcal{Q}_{ij}}$ and ${\mathcal{M}_{ij}}$ which determine the absence of gradient instabilities with high $\ell$, propagating in the angular direction, as well as tachyonic instabilities.
The components of matrix $\mathcal{M}_{ij}$ explicitly read:
{\small
\begin{subequations}
\label{eq:full_M}
\begin{align}
\label{eq:M11f}
\mathcal{M}_{11} =& \frac{\sqrt{AB}}{4 J^2} \frac{\mathcal{N}}{\cH^2}
+\frac{1}{4\ell(\ell+1)} \left( (\ell^2+\ell-2) \frac{\sqrt{B}}{\sqrt{A}}
\cH \cT^2 - \left[ \sqrt{B} \frac{\TT \cH \cT}{\cF}\right]'
\right)- \frac{\sqrt{A}}{4\sqrt{B}J^2 \; (\ell^2 +\ell -2)} \frac{\TT (\TT +\cW)}{\cF},\\
\label{eq:M12f}
\nonumber
\mathcal{M}_{12} =&  \frac{1}{2(\ell^2+\ell-2)}
\left(
\frac{\TT}{\cF J} \left[ \sqrt{AB} \; \cH (2 + J^2 \cD')\right]'
+ \frac{\sqrt{A}}{2\sqrt{B}J^2} 
\left( \ell(\ell+1) J \Big[ \cD \cW -  B \frac{\TT \cH}{\cF} 
\left(\frac{A'}{A} - 2 \frac{J'}{J}\right) \Big]
\right.\right.\\&\left.\left.
+ 2(\ell^2+\ell-2) B 
\left[\ell(\ell+1)J^2 \left(\cH \left(\frac{A'}{A} + 2\frac{J'}{J}\right) + \Gamma \pi'\right)
- (2\cH J J' + \Xi \pi') (2 + J^2 \cD')\right]\right)
\right),\\
\label{eq:M22f}
\nonumber
\mathcal{M}_{22} =& - \frac{\ell(\ell+1)}{(\ell^2+\ell-2)}
\left( \frac{\sqrt{AB}}{J^2} \cH \Big[\ell(\ell+1)(\ell^2+\ell-2) 
- (2+ J^2 \cD' -\ell(\ell+1))^2\Big] +\right.\\&\left.
\left[ \frac{\sqrt{AB}}{J} \frac{\cH}{\TT} \left( 2(\ell^2+\ell-2) B 
\left[\ell(\ell+1)J^2 \left(\cH \left(\frac{A'}{A} + 2\frac{J'}{J}\right) 
+ \Gamma \pi'\right)
- (2\cH J J' + \Xi \pi') (2 + J^2 \cD')\right]+ 
\right.\right.\right.\\&\left.\left.\left.
\phantom{\frac12} 
J \cD \big[\ell(\ell+1) \cW + \TT (2 + J^2 \cD')\big]
\right)
\right]'
\right) ,\nonumber
\end{align}
\end{subequations}
}
where
\[
\label{eq:calW}
\mathcal{W} = 3 B J J'(2 \cH J'  + \Gamma_1 J \pi') - 2 J (2\ell^2 + 2 \ell -1) (\cH - 2 F_4 B^2 {\pi'}^4),
\]
\[
\label{eq:calT}
\mathcal{T} = \sqrt{A} \left( \frac{2\cH JJ' + \Xi \pi'}{\cH J}
+ \frac{J}{(\ell^2+\ell-2)} \left[ \frac{\TT}{\cF J}\right]'\right),
\]
\[
\label{eq:calD}
\mathcal{D} = \frac{[\cH^2 B J^2]'}{J^2 \cH \cF}.
\]
We do not give here an explicit expression for $\det{\mathcal{M}}$, since it does not have a concise and clear form.

As for the matrix ${\mathcal{Q}_{ij}}$, 
for the sake of simplicity we transform it to the off-diagonal form, so that the terms $\mathcal{Q}_{ii} v^i {v^i}'$ are included into $\mathcal{M}_{ii}$ upon integration by parts and the only non-trivial component reads:
\[
\label{eq:Q12f}
\begin{aligned}
\mathcal{Q}_{12} =& 
2 \left[ B^{3/2} J \frac{\cH^2}{\cF}\right]' \mathcal{T} 
+ B^{3/2} J \frac{\cH^2}{\cF} \mathcal{T}' 
-\frac{\sqrt{AB}}{2 J (\ell^2 + \ell -2)} \frac{\cH}{\cF}
\Big(\ell(\ell+1) \cW + \TT (2 + J^2 \cD')
\Big) \\
&- 4 \ell(\ell+1) BJ \frac{\cH}{\TT} \mathcal{M}_{11}.
\end{aligned}
\]

Let us find the constraints coming from matrices $\mathcal{M}_{ij}$ and $\mathcal{Q}_{ij}$ together as soon as one requires the energy density to be positive and the dispersion relation to be regular for all momenta~\cite{RubakovNEC}. 
We rewrite the last three terms from the action~\eqref{eq:action_KGQM_new} in the following form:
\[
\label{eq:MQform}\frac12 {\mathcal{G}}_{ij} v^{i\prime} v^{j\prime} + {\mathcal{Q}}_{ij} v^i v^{j\prime} + \frac12 {\mathcal{M}}_{ij} v^i v^j =
\begin{pmatrix}
{v^2}' \\ {v^1}' \\ {v^2} \\ {v^1}
\end{pmatrix}^T
\begin{pNiceArray}{cc|cc}
  \mathcal{G}_{22} & \mathcal{G}_{12} & 0 & \mathcal{Q}_{12}/2 \\
  \mathcal{G}_{12} & \mathcal{G}_{11} & 0 & 0\\
  \hline
  0 & 0 & \mathcal{M}_{22} & \mathcal{M}_{12} \\
  \mathcal{Q}_{12}/2 &0 & \mathcal{M}_{12} & \mathcal{M}_{11} 
\end{pNiceArray}
\begin{pmatrix}
{v^2}' \\ {v^1}' \\ {v^2} \\ {v^1}
\end{pmatrix},
\]
{which is the spatial part of the energy density.}
We suggest that 
${v^i}$ and ${v^i}'$ are independent variables which corresponds to the most general case possible in this setting and gives a sufficient stability condition which, however, might be not the necessary one. 
As before we require that the block matrix in eq.~\eqref{eq:MQform} is positive-definite, which is the case provided that:
\begin{subequations}
\label{eq:MQconditions}	
\begin{align}
\label{eq:Gagain}	
& \mathcal{G}_{11} > 0, \quad \det\mathcal{G} > 0, \\
\label{eq:M22constr}
& \mathcal{M}_{22} > 0, \\
\label{eq:MQconstr}
& \det\mathcal{G} \cdot \det\mathcal{M} > \frac{\mathcal{Q}_{12}^2}{4} \mathcal{M}_{22} \cdot \mathcal{G}_{11}.
\end{align}
\end{subequations}
We see that the inequalities~\eqref{eq:Gagain} naturally coincide with eq.~\eqref{eq:noradial}, which ensures the absence of radial gradient instabilities. The rest of constraints in eqs.~\eqref{eq:M22constr} and~\eqref{eq:MQconstr} are sufficient to avoid both gradient instabilities propagating in the angular direction and tachyons. Unlike the cases of ghost and radial gradient instabilities above, the stability criteria~\eqref{eq:M22constr} and~\eqref{eq:MQconstr} do not give a concise constraints for the Lagrangian functions similar to eqs.~\eqref{eq:noghost_explicit} and~\eqref{eq:noradial_explicit}. 

However, if we focus on modes with high multipole $\ell \gg 2$ only,
the corresponding stability conditions~\eqref{eq:M22constr} and~\eqref{eq:MQconstr}
get simplified. Indeed, for {high energy modes} the main source of constraints are those parts of matrices $\mathcal{Q}_{ij}$ and $\mathcal{M}_{ij}$
which are proportional to $\ell^2$.
It follows from eq.~\eqref{eq:Q12f} that in the leading order $\mathcal{Q}_{12}$ does not depend on $\ell$
and, hence, does not provide any constraints related to angular gradient instabilities. Thus, $\mathcal{Q}_{12}$ does not contribute to the inequality~\eqref{eq:MQconstr} in the limit of $\ell \gg 2$. Moreover, since $\det \mathcal{M} \propto \ell^4$ and $\mathcal{M}_{22} \propto \ell^2$  the inequality~\eqref{eq:MQconstr} reduces to 
$\det \mathcal{M} > 0$
in the limit of $\ell \gg 2$, provided that $\det \mathcal{G} > 0$ due to no radial gradient instability condition~\eqref{eq:noradial} or~\eqref{eq:Gagain}. Hence, together with eq.~\eqref{eq:M22constr} the set of conditions ensuring the absence of angular gradient instabilities for high energy modes amounts to positive definiteness of the matrix ${\mathcal{M}}^{(\ell^2)}_{ij}$
\footnote{Here we swap $\mathcal{M}_{22}$ for $\mathcal{M}_{11}$ for the sake of simplicity.}:
\[
\label{eq:noangular}
\mathcal{M}^{(\ell^2)}_{11} > 0, \qquad \det \mathcal{M}^{(\ell^2)} > 0,
\]
where the superscript $\ell^2$ emphasizes that this is only
a certain part of matrix ${\mathcal{M}}_{ij}$ that is proportional to $\ell^2$ (i.e. the angular part of the Laplace operator), and
\begin{equation}
\label{eq:M11_l2}
{\mathcal{M}}^{(\ell^2)}_{11} = \ell^2  \dfrac{\sqrt{A}}{\sqrt{B}}\dfrac{(\mathcal{H} - 2 F_4 B^2 \pi'^4)^2}{\mathcal{F}},
\end{equation}
{
\begin{equation}
\label{eq:detM_l2}
\det{\mathcal{M}}^{(\ell^2)} = - \ell^4 \dfrac{ B \left( \mathcal{H} - 2 F_4 B^2 \pi'^4\right)^2 \mathcal{H}^2}{4 J^2 \;\mathcal{F}^2 } 
\left(J^4 \frac{\sqrt{A}}{\sqrt{B}}
\left[ \frac{\sqrt{B}}{\sqrt{A} J^3
 } \mathcal{P}_4 \right]'  +\frac{2 A \cG}{B \cH^2} +
 \frac{(\cF \mathcal{P}_4 - (A'J - 2 A J'))^2}{4 A  \cF}
\right),
\end{equation}
}
with
\begin{equation}
\mathcal{P}_4 = \dfrac{\mathcal{H} A' J + 2 \mathcal{G} A J' + \Gamma A J \pi'}{\mathcal{H} \left(\mathcal{H} - 2 F_4 B^2 \pi'^4\right)},
\end{equation}
and the rest of ${\mathcal{M}}^{(\ell^2)}_{ij}$ are given in Appendix B for completeness.

It immediately follows from eq.~\eqref{eq:M11_l2} that ${\mathcal{M}}^{(\ell^2)}_{11}$ is always non-negative (provided that eq.~\eqref{eq:stability_F} holds in the odd-parity sector), while the positivity of $\det{\mathcal{M}}^{(\ell^2)}$ introduces an additional constraint for the model:
\begin{equation}
\label{eq:noangular_explicit}
J^4 \frac{\sqrt{A}}{\sqrt{B}} \left[ \frac{\sqrt{B}}{\sqrt{A} J^3
} \mathcal{P}_4 \right]' < - 
\left(\frac{2 A \cG}{B \cH^2} +
 \frac{(\cF \mathcal{P}_4 - (A'J - 2 A J'))^2}{4 A  \cF} \right).
\end{equation}

As for additional (sub)luminality constraints, in full analogy with the radial sound speeds~\eqref{eq:speed_even_radial} matrices ${\mathcal{M}}^{(\ell^2)}_{ij}$ and $\mathcal{K}_{ij}$ define the angular sound speeds $c^2_{a1,a2}$,
which are given 
by the 
eigenvalues of matrix $(A^{-1}J^2)({\mathcal{K}})^{-1}{\mathcal{M}}^{(\ell^2)}$. The corresponding system of equations for $c^2_{a1,a2}$ reads: 
\begin{subequations}
\label{eq:angular_speed}
\begin{align}
\label{eq:product}
c^2_{a1} \cdot c^2_{a2} &= -  \frac{\ell^4 \; A}{J^6 (2 \mathcal{P}_1 - \cF)} 
\xi^2 \left(J^4 \frac{\sqrt{A}}{\sqrt{B}}
\left[ \frac{\sqrt{B}}{\sqrt{A} J^3
 } \mathcal{P}_4 \right]' + \frac{2 A \cG}{B \cH^2} 
+  
 \frac{\left(\cF \mathcal{P}_4 - (A'J - 2 A J')\right)^2}{4 A \cF}
\right)
,\\
c^2_{a1} + c^2_{a2} &= -  \frac{\ell^2 }{J^4 (2 \mathcal{P}_1 - \cF)} 
 \left[\xi^2 \left(J^4 \frac{\sqrt{A}}{\sqrt{B}}
\left[ \frac{\sqrt{B}}{\sqrt{A} J^3
 } \mathcal{P}_4 \right]' + \frac{2 A \cG}{B \cH^2}\right) +
\xi J \left(\cF \mathcal{P}_4 - (A'J - 2 A J')\right) - 2 A\;J^2 \mathcal{P}_1
\right].
\end{align}
\end{subequations}
The explicit expressions for $c^2_{a1,a2}$ again are not really illuminating. However, the system~\eqref{eq:angular_speed} still enables one to assess if superluminalities are present for any specific model~\eqref{eq:lagrangian}. Let us also note that the constraint~\eqref{eq:noangular_explicit} coming from positivity of $\det{\mathcal{M}}^{(\ell^2)}$ ensures that speeds $c^2_{a1,a2}$ are real, see eq.~\eqref{eq:product}.  

Thus, the full set of stability constraints in the even-parity sector consists of no ghost condition~\eqref{eq:noghost_explicit} and a group of conditions~\eqref{eq:MQconditions}, which govern gradient instabilities and tachyons. In the case of high energy modes the angular gradient instabilities 
can be constrained by inequalities~\eqref{eq:noangular} or explicitly~\eqref{eq:noangular_explicit}.

\section{Conclusion}
\label{sec:conclusion}

In this note we have considered linear perturbations about an arbitraty static, spherically symmetric background in the quadratic beyond Horndeski theory. We have focused on analysing the behaviour of the even-parity modes, since the analysis of slow tachyonic modes was still missing in the previous works. In this work we have derived the quadratic action for perturbations in the parity even sector in full generality and explicitly presented the matrices $\mathcal{Q}$ and $\mathcal{M}$, featuring lower derivative terms including mass terms for the Regge-Wheeler-unitary gauge. In result, we have formulated a complete set of stability criteria for both odd-parity and even-parity sector constraining ghosts, gradient instabilities and tachyons. We have also studied the constraints on the sound speeds in both radial and angular propagation directions, which enables one to eliminate superluminal modes in both parity sectors.
Let us note
that this complete set of constraints is sufficient to claim stability, but might be 
{not a necessary one}.

The main advantage of the formulated set of stability conditions is its generality as it addresses the stability issue for an arbitrary background in a vast class of theories. In particular, the subset of these stability conditions were utilized for construction of a stable wormhole solution in e.g. Refs.~\cite{wormRMV1,wormRMV2}. 
{Let us note in passing that both solutions~\cite{wormRMV1,wormRMV2}, do not satisfy the complete set of stability conditions and, hence, cannot claim to be fully stable.} 
We expect that the new complete set of stability condition will enable one to make further development in exploring various stable solutions featuring compact objects like black holes and wormholes.

\section*{Acknowledgements}
The work
has been supported by the Foundation for the Advancement of Theoretical Physics and
Mathematics “BASIS”.

\section*{Appendix A}
In this Appendix we gather the explicit form of coefficients
$a_i$, $b_i$, $c_i$, $d_i$, $e_i$ and $p_i$ entering the quadratic action in
a gauge invariant form~\eqref{eq:action_even}. The coefficients 
are given in terms of combinations $\cH$, $\cF$, $\cG$, $\Xi$, $\Gamma$
and $\Sigma$ introduced in the main body of the text in eqs.~\eqref{eq:cal_FGH},~\eqref{eq:Xi} and~\eqref{eq:GammaSigma}. The coefficients below also involve background Einstein and scalar field equations of motion denoted as
\[
\mathcal{E}_A = 0, \qquad \mathcal{E}_B = 0,
\qquad \mathcal{E}_J = 0, \qquad \mathcal{E}_{\pi} = 0.
\] 
We do not use them in current work so we do not give them here, but their explicit expressions can be read off in Appendix A in Ref.~\cite{wormRMV2}.
\begin{eqnarray}
a_1&=&\sqrt{AB}\,\Xi,\\
a_2&=&\frac{\sqrt{AB}}{2\pi'}\left[
2\pi'\Xi'-\left(2\pi''-\frac{A'}{A}\pi'\right)\Xi
+2JJ'\left(\frac{A'}{A}-\frac{B'}{B}\right){\cal H}-4{\cal H}JJ''
\right.\\ &&\left.
+\frac{2J^2}{B}\left({\cal E}_B-{\cal E}_A\right) \right] - 
 2  \sqrt{AB^3} \pi'^3 \cdot F_4 ,\\
a_3&=&-\frac{\sqrt{AB}}{2}\left(\pi'\Xi+2JJ'{\cal H}\right),\\
a_4&=&\sqrt{AB}\,{\cal H},\\
a_5&=&-\sqrt{\frac{A}{B}}J^2\frac{\partial{\cal E}_A}{\partial\pi }=a_2'-a_1'',\\
a_6&=&-\sqrt{\frac{A}{B}} \frac{1}{J\pi'} \left( J {\cal H}'+J'{\cal H}-J'{\cal F} \right) + \sqrt{AB} \pi'^2 \left( B' \pi' + 2 B \pi''\right) \cdot F_4, \\
a_7&=&a_3'+\frac{J^2}{2} \sqrt{\frac{A}{B}} {\cal E}_B,\\
a_8&=&-\frac{a_4}{2B} + \sqrt{AB^3} \pi'^4 \cdot F_4, \\
a_9&=&\frac{\sqrt{A}}{J}\frac{\D}{\D r}\left(
J\sqrt{B}{\cal H}
\right), \\
a_{10}&=&-\frac{1}{2} a_4,
\\
a_{11}&=&-\frac{1}{2}\sqrt{{A}{B}}\left[\cH' + \frac12 \left(\frac{B'}{B}- 2\frac{J'}{J} \right)\cH \right],
\\
a_{12}&=& \sqrt{{A}{B}}\left[\frac{J'}{J} \cH' + \frac12 \left(\frac{B'}{B} \frac{J'}{J} + 2\frac{J''}{J} \right)\cH \right],
\\
a_{13}&=& J^2\; a_4,
\\
a_{14}&=& \sqrt{{A}{B}} \;J^2 \left[\cH' + \frac12 \left(\frac{B'}{B} +6\frac{J'}{J} \right)\cH \right],
\\
a_{15} &=& \sqrt{\frac{A}{B}}\; \cF,
\\
a_{16} &=& -\frac12 a_{15},
\\
b_1&=&\frac{1}{2}\sqrt{\frac{B}{A}}{\cal H},
\end{eqnarray}
\begin{eqnarray}
b_2&=&-2\sqrt{\frac{B}{A}}\Xi,
\\
b_3&=&\sqrt{\frac{B}{A}}\frac{1}{\pi'}
\left[
\left(2\pi''+\frac{B'}{B}\pi'\right)\Xi -2JJ'\left(\frac{A'}{A}-\frac{B'}{B}\right){\cal H}
+\frac{2J^2}{B}{\cal E}_A+4JJ''{\cal H}
\right]
,
\\
b_4&=&\sqrt{\frac{B}{A}}\left(\pi'\Xi+2JJ'{\cal H}\right),
\\
b_5&=&-2b_1,\\
b_{6} &=& \sqrt{\frac{B}{A}}\; \cH,\\
b_{7} &=& -\frac12 \sqrt{\frac{B}{A}} \left(\frac{A'}{A} + 2 \frac{J'}{J} \right) \cH,
\\
b_{8} &=& - 2\sqrt{\frac{B}{A}} J^2 \cH,
\\
b_{9} &=& \sqrt{\frac{B}{A}} J^2 \left(\frac{A'}{A} - 2 \frac{J'}{J}  \right) \cH,
\\
b_{10} &=& -b_6,
\\
b_{11} &=& \sqrt{\frac{B}{A}} \frac{A'}{A} \cH,
\\
c_1&=&-\frac{1}{\sqrt{AB}}\Xi - 4\sqrt{\frac{ B^3}{A}} J J' \pi'^3 \cdot F_4,
\\
c_2&=&-\sqrt{AB} \left( \frac{A'}{2A} \Xi+JJ'\Gamma-\frac{J^2 \pi'}{X} \Sigma \right)
,
\\
c_3&=&J^2\sqrt{\frac{A}{B}} \frac{\partial {\cal E}_B}{\partial \pi},\\
c_4&=&\frac{1}{2}\sqrt{\frac{A}{B}} \Gamma + \sqrt{\frac{ B^3}{A}} \left(A' J + 2A J'\right) \pi'^3
 \cdot F_4,
\\
c_5&=&
-\frac{1}{2}\sqrt{AB}\left(\pi'\Gamma+\frac{A'}{A}{\cal H}
+\frac{2J'}{J}{\cal G}\right),
\\
c_6&=&\frac{J^2}{2} \sqrt{\frac{A}{B}} \left( \Sigma+\frac{A'B \pi'}{2J^2 A}\Xi+\frac{B\pi'J'}{J}\Gamma-\frac{1}{2} {\cal E}_B+\frac{BJ'^2}{J^2} {\cal G}+\frac{A'BJ'}{JA}{\cal H} \right),
\\
c_{7} &=& \frac12 \frac1{\sqrt{{A}{B}}} (\cH - 2 F_4 B^2 \pi'^4),
\\
c_{8} &=& -2\; J^2\; c_7,
\\
c_{9} &=& \frac14 \sqrt{{A}{B}} \left[\Gamma \pi' + \left(\frac{A'}{A} + 2 \frac{J'}{J} \right)\cH \right],
\\
c_{10} &=& - \frac14\sqrt{{A}{B}} \left[
\frac{J'}{J} \left( \Gamma + \frac{J'}{J} \Gamma_2\right)\pi' - \frac14\frac{A'}{A}\left( \Gamma_1 - 2 \frac{J'}{J} \Gamma_2\right)\pi'+
\right.\\\nonumber&&\left.
\cH \frac{J'}{J} \left(\frac{A'}{A} + 3 \frac{J'}{J} \right)+
\frac{1}{BJ^2} (\cH - 2 F_4 B^2 \pi'^4) \right],
\\
c_{11} &=& -2 J^2 c_9,
\\
c_{12} &=& - \frac12 \sqrt{{A}{B}} J^2 \left[
\frac{1}{2 J^2}\left(\frac{A'}{A} + \frac{J'}{J} \right)\Xi\pi' +
\frac34 \frac{J'}{J} \left( \Gamma_1 - 2 \frac{J'}{J} \Gamma_2\right)\pi' 
\right.\\\nonumber&&\left.
+ \frac{J'}{J} \left(\frac{A'}{A} + \frac{J'}{J} \right) \cH +
\frac{1}{BJ^2} (\cH - 2 F_4 B^2 \pi'^4) \right],
\\
c_{13} &=& A \;c_7,
\\
c_{14} &=& 2\; c_7,
\end{eqnarray}
\begin{eqnarray}
d_1&=&b_1,
\\
d_2&=&\sqrt{AB}\,\Gamma,
\\
d_3&=& \frac{\sqrt{AB}}{J^2}\left[
\frac{2JJ'}{\pi'}\left(\frac{A'}{A}-\frac{B'}{B}\right){\cal H}
-J^2\left(\frac{2J'}{J}-\frac{A'}{A}\right)\frac{\partial{\cal H}}{\partial\pi}
+\frac{2}{B\pi'}\left({\cal F}-{\cal G}\right)\right.
\nonumber\\&&\left.
-\frac{J^2}{2\pi'}\left(2\pi''+\frac{B'}{B}\pi'\right)
\left(\Gamma_1+\frac{2J'}{J}\Gamma_2\right)
-\frac{2J^2}{B\pi'}({\cal E}_A-{\cal E}_B)-\frac{4JJ''}{\pi'}{\cal H}
\right],
\\
d_4&=&\frac{\sqrt{AB}}{J^2}
\left({\cal G}-J^2{\cal E}_B\right),
\\
d_{5} &=& - \frac{\sqrt{{A}{B}}}{J^2} \cH,
\\
d_{6} &=& 2\frac{\sqrt{{A}{B}}J'}{J^3} \cH,
\\
d_{7} &=& a_4,
\\
d_{8} &=& b_6,
\\
d_{9} &=& -2 \sqrt{\frac{B}{A}}\frac{J'}{J} \cH,
\\
e_1&=&\frac{1}{2\sqrt{AB}}\left[
\frac{J^2}{X}({\cal E}_A-{\cal E}_B)-\frac{2}{\pi'}\Xi'+\left(\frac{A'}{A}
-\frac{X'}{X}\right)\frac{\Xi}{\pi'}
+\frac{2BJ'^2}{X}{\cal F}-\frac{2JJ'B}{X}{\cal H}' \right.\\&&\left.
-{\cal H}\frac{B^2J'^2}{JXA}\frac{\D}{\D r}\left(\frac{J^2A}{B}\right)+\frac{2BJJ''}{X}{\cal H}
\right] - \frac{4}{B \pi'^2} \cdot \frac{\mbox{d}}{\mbox{d} r}
\left[\sqrt{\frac{B^5}{A}} J J' \pi'^4 \cdot F_4\right], \nonumber \\
e_2&=&-\sqrt{AB}\frac{J^2}{X}\Sigma,\\
e_3&=&J^2 \sqrt{\frac{A}{B}} \frac{\partial {\cal E}_\pi}{\partial \pi},
\\
e_4&=&\frac{\sqrt{AB}J'^2}{8X}\left(-\frac{4{\cal G}}{J^2}-\frac{4({\cal E}_A-{\cal E}_B)}{BJ'^2}+\frac{2A'{\cal H}'}{AJ'^2}+
\frac{4{\cal G}'}{JJ'}+\frac{4}{BJ^2J'^2}\left(1-\frac{JJ'BA'}{A}\right){\cal F}
\nonumber\right.\\ &&\left.-
\frac{4{\cal H}}{BJ^2J'^2}\left(1-BJ'^2(1+\frac{2A'J}{AJ'})+J(B'J'+2BJ'')\right)-\frac{2\pi'}{J'^2}\Gamma'
\nonumber\right.\\ &&\left.
+\frac{2\pi'}{JJ'}\left(-2+\frac{A'J}{AJ'}\right)\frac{\partial{\cal H}}{\partial\pi}
+\frac{\Xi\pi'}{J^3J'}\left(2-\frac{A'J}{AJ'}\right)\left[\frac{A'BJJ'}{A}-2+2BJ'^2\right]
\nonumber\right.\\ &&\left.
+\frac{\Gamma_1\pi'}{2J'}\left[\frac{4}{J}+\frac{A'^2BJ}{A^2}-\frac{4A'}{AJ'}-\frac{4BJ'^2}{J}
-\frac{2B'}{BJ'}+\frac{2BJ''}{BJ'^2}-\frac{4\pi''}{\pi'J'}
\right]
-\nonumber\right.\\ &&\left.
-
\frac{\Gamma_2\pi'}{J'}\left[\frac{2A'}{AJ}+\frac{A'^2}{A^2J'}(1-BJ'^2)-\frac{4J'}{J^2}(1-BJ'^2)+\frac{2B'}{BJ}+\frac{4\pi''}{\pi'J}
\right]\right) - e_4^{BH},
\\
e_4^{BH}&=& -\frac{\pi'^2}{J^2 J'} \sqrt{\frac{B^3}{A}}
\left(A'' J^2 J' - 4 A J'^3 +A' J(J'^2 - J J'')\right) \cdot F_4
\\\nonumber&&
- \frac{A'J + 2AJ'}{B J^2 J' \pi'^2} \cdot
\frac{\mbox{d}}{\mbox{d} r} \left[\sqrt{\frac{B^5}{A}} J J' \pi'^4 \cdot F_4\right],
\\
e_{5} &=& - \frac{1}{\sqrt{{A}{B}}} (2F_4 B^2 \pi'^3),
\\
e_{6} &=& \frac{1}{\sqrt{{A}{B}}\pi'} \left[
(\cH - 2 F_4 B^2 \pi'^4)' + \frac{J'}{J} (\cH - \cF) + F_4 B^2 \pi'^4 \frac{A'}{A}
\right],
\\
e_{7} &=&  2\frac{ J^2}{\sqrt{{A}{B}}} (2F_4 B^2 \pi'^3),
\\
e_{8} &=&-2 J^2 \frac{1}{\sqrt{{A}{B}}\pi'} \left[
(\cH - 2 F_4 B^2 \pi'^4)' + \frac{J'}{J} (\cH - \cF) + F_4 B^2 \pi'^4 \left(\frac{A'}{A} -4 \frac{J'}{J}\right)
\right],
\end{eqnarray}
\begin{eqnarray}
e_{9} &=&  2\frac{1}{\sqrt{{A}{B}}} (2F_4 B^2 \pi'^3),
\\
e_{10} &=& = -2 e_6,
\\
e_{11} &=& - \frac12 {\sqrt{{A}{B}}}\; \Gamma,
\\
e_{12} &=& -\frac12 \frac{\sqrt{{A}{B}}}{\pi'}
\left[
2\cH \left(\frac{B'}{B}\frac{J'}{J} +\frac{J'^2}{J^2} + 2 \frac{J''}{J}\right) + \frac{\Xi}{B J^2} (B'\pi' + 2 B \pi'') - \Gamma_1 \frac{J'}{J} \pi'+
\right.\\\nonumber&&\left.
\frac12 \pi' \left( \left(\frac{A'}{A} +2 \frac{J'}{J}\right) \left(\Gamma-\Gamma_2 \left(\frac{A'}{A} +2 \frac{J'}{J}\right) + 2 \frac{\partial\cH}{\partial\pi}
\right) - 4 \frac{\partial\cH}{\partial\pi} \frac{A'}{A}
\right) - \frac{2}{B J^2} (\cF - 2 F_4 B^2 \pi'^4)
\right],
\\
e_{13} &=& \sqrt{{A}{B}} J^2 \Gamma,
\\
e_{14} &=& \frac{\sqrt{{A}{B}}}{\pi'}
\left[
2\cH \left(\frac{B'}{B}\frac{J'}{J} +\frac{J'^2}{J^2} + 2 \frac{J''}{J}\right) + \frac{\Xi}{B J^2} ( B'\pi' + 2  B \pi'') +(2\Gamma - \Gamma_1) \frac{J'}{J} \pi'+
\right.\\\nonumber&&\left.
\frac12 \pi' \left( \left(\frac{A'}{A} +2 \frac{J'}{J}\right) \left(\Gamma-\Gamma_2 \left(\frac{A'}{A} +2 \frac{J'}{J}\right) + 2 \frac{\partial\cH}{\partial\pi}
\right) - 4 \frac{\partial\cH}{\partial\pi} \frac{A'}{A}
\right)
\right.\\\nonumber&&\left.
 - \frac{2}{B J^2} (\cF - \ell(\ell+1) F_4 B^2 \pi'^4)
\right],
\\
e_{15} &=& \sqrt{{A}{B}} (2F_4 B \pi'^3),
\\
e_{16} &=& - \frac{\sqrt{A}}{\sqrt{B}} \left(
\frac{1}{J^2 \pi'} \left[ \cH' + \frac{A'}{2A} (\cH -\cF) - F_4 B{\pi'}^3 
(B'\pi' +2B\pi'')\right]
+ \frac{\partial \mathcal{E}_C}{\partial \pi} 
\right.\\\nonumber&&\left.
- \mathcal{E}_A \left[ \frac{1}{2\pi'} \left(\frac{A'}{A} +2 \frac{J'}{J}\right) + \frac{1}{2B{\pi'}^2} (B'\pi'+2B \pi'') \right]
\right),
\\
e_{17} &=&2 J^2 \frac{\sqrt{A}}{\sqrt{B}}
\left( \frac{\partial \mathcal{E}_C}{\partial \pi} 
- \mathcal{E}_A \left[ \frac{1}{2\pi'} \left(\frac{A'}{A} +2 \frac{J'}{J}\right) + \frac{1}{2B{\pi'}^2} (B'\pi'+2B \pi'') \right]
\right),
\\
e_{18} &=& \frac{\sqrt{A}}{\sqrt{B}}\frac{1}{\pi'} \left( \cH' + \frac{A'}{2A} (\cH -\cF) - F_4 B{\pi'}^3 
(B'\pi' +2B\pi'')\right),
\\
p_{1} &=& \frac12 b_6,
\\
p_{2} &=& \frac12 \sqrt{\frac{B}{A}} \left[
2 \frac{J'}{J} \cH' + \cH \left(
-\frac{A'}{A} \frac{J'}{J} + \frac{B'}{B} \frac{J'}{J} + 2 \frac{J'^2}{J^2} + 2 \frac{J''}{J}
\right) - \frac{2}{B J^2} \cF
\right],
\\
p_{3} &=& \frac{1}{\sqrt{{A}{B}}J^2} \cF,
\\
p_{4} &=&  - \frac{1}{\sqrt{{A}{B}}} \cF,
\\
p_{5} &=& -\frac14 p_3,
\\
p_{6} &=& \frac{\sqrt{{A}{B}}}{4 J^2} \cH,
\\
p_{7} &=& \frac{1}{4 J^2} \sqrt{{A}{B}} \left[
2 \frac{J'}{J} \cH' + \cH \left(
\frac{A'}{A} \frac{J'}{J} + \frac{B'}{B} \frac{J'}{J} - 2 \frac{J'^2}{J^2} + 2 \frac{J''}{J}
\right)
\right],
\\
p_{8} &=& -\frac{J^2}{2 \sqrt{{A}{B}}} \cF,
\\
p_{9} &=& \frac12 \sqrt{{A}{B}} J^2 \cH,
\\
p_{10} &=& \frac{1}{2\sqrt{{A}{B}}} \cF,
\\
p_{11} &=& - \frac12 a_4,
\\
p_{12} &=& {\sqrt{{A}{B}}} \frac{J'}{J} \cH.
\end{eqnarray}

\section*{Appendix B}
Here we put the rest of components of matrices ${\mathcal{K}}_{ij}$, ${\mathcal{G}_{ij}}$ and ${\mathcal{M}}^{(\ell^2)}_{ij}$ entering the quadratic
action~\eqref{eq:action_KGQM_new} for completeness:
\begin{equation}
{\mathcal{K}}_{12} = {\mathcal{K}}_{21} = \frac{\sqrt{B}}{2(\ell+2)(\ell-1) \sqrt{A}} \left[2(\ell+2)(\ell-1) (2\cH J J' + \Xi \pi') -
\frac{\left[\cH^2 B J^2 \right]'}{\cH B J} \frac{\TT}{\cF} \right],
\end{equation}
\begin{equation}
{\mathcal{K}}_{22} =  \frac{4 \ell(\ell+1) \sqrt{A} \sqrt{B}}{\TT^2} \left[\ell(\ell+1) \frac{B}{A} (2 \mathcal{P}_1 -\cF) (2\cH J J' + \Xi \pi')^2
+ (\ell+2)(\ell-1) \cF\; {\mathcal{K}}_{12}^2 \right],
\end{equation}

\begin{equation}
{\mathcal{G}}_{12} = \mathcal{G}_{21} = AB\frac{\cG}{\cF} {\mathcal{K}}_{12}
\end{equation}

\begin{equation*}
{\mathcal{G}}_{22} = \frac{4\ell(\ell+1) \sqrt{A} \sqrt{B}}{\TT^2} \frac{\cG}{\cH} \left[ \ell(\ell+1) B^2 (2 J^2 \Gamma \mathcal{H} \Xi \pi'^2  - \mathcal{G} \Xi^2 \pi'^2-4 J^4 \Sigma \mathcal{H}^2/B )
+ \frac{(\ell+2)(\ell-1)\cF^2}{AB \;\cG} {\mathcal{G}}_{12}^2\right],
\end{equation*}

\begin{equation}
{\mathcal{M}}^{(\ell^2)}_{12} =  {\mathcal{M}}^{(\ell^2)}_{21} = - \ell^2  \frac{ \cH (\cH - 2 F_4 B^2 \pi'^2) } {2 \cF J} \frac{\sqrt{A}}{\sqrt{B}} \left[ 2 \frac{\left[\cH^2 B J^2 \right]'}{\cH^2 J} + \frac{B(A'J - 2 A J')}{A} - \frac{B}{A} \cF \mathcal{P}_4
 \right],
\end{equation}

\begin{multline}
 {\mathcal{M}}^{(\ell^2)}_{22} = -\ell^2 \;\frac{ \cH^2}{\cF} \frac{\sqrt{A} \sqrt{B}}{J^2} \left[ 2\frac{ \cF \cG }{\cH^2} + \frac{\sqrt{B}}{\sqrt{A}} J^4\; \cF 
\left[\frac{\sqrt{B}}{\sqrt{A}J^3} \mathcal{P}_4\right]' +\frac{1}{B} \left(\frac{\left[\cH^2 B J^2 \right]'}{\cH^2 J}\right)^2
 \right] \\
  - \frac{2}{(\cH - 2 F_4 B^2 \pi'^2)} \frac{\left[\cH^2 B J^2 \right]'}{\cH J^2} {\mathcal{M}}_{12}^{(\ell^2)}.
\end{multline}



\end{document}